\documentclass[floats,aps,showpacs,twocolumn,superscriptaddress]{revtex4-1}

\usepackage[dvips]{graphicx}
\usepackage{amsfonts}
\usepackage{amsmath,amssymb}
\usepackage{dsfont}
\usepackage{placeins}
\usepackage{afterpage}
\usepackage{flafter}
\usepackage[usenames,dvipsnames]{color}

\newcommand{\heff}{h}
\newcommand{\myi}{{\text{i}}}

\newcommand{\ireg}{m}
\newcommand{\ich}{\text{ch}}

\newcommand{\paths}{\nu}
\newcommand{\maslov}{\phi}

\newcommand{\icurve}{\mathcal{C}}

\newcommand{\fcurve}{\mathcal{C}}

\newcommand{\act}{I}
\newcommand{\ang}{\theta}

\newcommand{\Ii}{I_{\ireg}}
\newcommand{\If}{I_{\ich}}
\newcommand{\Ib}{I_{\text{b}}}

\newcommand{\Eb}{E_{\text{b}}}
\newcommand{\Ef}{E'}

\newcommand{\qi}{q}
\newcommand{\qf}{q'}

\newcommand{\pin}{p}
\newcommand{\pf}{p'}

\newcommand{\qint}{q}
\newcommand{\pint}{p}

\newcommand{\map}{U}
\newcommand{\qmap}{\widehat{U}}
\newcommand{\Hreg}{H_{\text{reg}}}
\newcommand{\Sif}{\mathcal{S}}
\newcommand{\Sifind}[1]{\Sif_{#1}}

\newcommand{\Smapind}[1]{\mathcal{S}_{#1}^{\map}}

\newcommand{\Si}{\int\limits_{\icurve_{\ireg,\paths}} \hspace*{-0.25cm}
\pint (\qint)\,\text{d} \qint}
\newcommand{\Sf}{\int\limits_{\fcurve_{\ich,\paths}} \hspace*{-0.25cm}
\pint (\qint)\,\text{d} \qint}

\newcommand{\myexp}[1]{\exp{\left(#1\right)}}

\newcommand{\myIm}[1]{\text{Im}{#1}}
\newcommand{\myRe}[1]{\text{Re}{#1}}

\begin{document}

\title{Complex paths for regular-to-chaotic tunneling rates}

\author{Normann Mertig}
\affiliation{Institut f\"ur Theoretische Physik, Technische Universit\"at
             Dresden, 01062 Dresden, Germany}
\affiliation{Max-Planck-Institut f\"ur Physik komplexer Systeme, N\"othnitzer
Stra\ss{}e 38, 01187 Dresden, Germany}

\author{Steffen L\"ock}
\affiliation{Institut f\"ur Theoretische Physik, Technische Universit\"at
             Dresden, 01062 Dresden, Germany}
\affiliation{Max-Planck-Institut f\"ur Physik komplexer Systeme, N\"othnitzer
Stra\ss{}e 38, 01187 Dresden, Germany}

\author{Arnd B\"acker}\
\affiliation{Institut f\"ur Theoretische Physik, Technische Universit\"at
             Dresden, 01062 Dresden, Germany}
\affiliation{Max-Planck-Institut f\"ur Physik komplexer Systeme, N\"othnitzer
Stra\ss{}e 38, 01187 Dresden, Germany}

\author{Roland Ketzmerick}
\affiliation{Institut f\"ur Theoretische Physik, Technische Universit\"at
             Dresden, 01062 Dresden, Germany}
\affiliation{Max-Planck-Institut f\"ur Physik komplexer Systeme, N\"othnitzer
Stra\ss{}e 38, 01187 Dresden, Germany}

\author{Akira Shudo}
\affiliation{Max-Planck-Institut f\"ur Physik komplexer Systeme,
N\"othnitzer
Stra\ss{}e 38, 01187 Dresden, Germany}
\affiliation{\mbox{Department of Physics, Tokyo Metropolitan University,
Minami-Osawa, Hachioji, Tokyo 192-0397, Japan}}

\date{\today}

\begin{abstract}

In generic Hamiltonian systems tori of regular motion are dynamically separated
from regions of chaotic motion in phase space. Quantum mechanically these
phase-space regions are coupled by dynamical tunneling. We introduce a
semiclassical approach based on complex paths for the prediction of dynamical
tunneling rates from regular tori to the chaotic region. This approach is
demonstrated for the standard map giving excellent agreement
with numerically determined tunneling rates.

\end{abstract}
\pacs{05.45.Mt, 03.65.Sq}

\maketitle
\noindent


Tunneling is a fundamental manifestation of quantum mechanics. Its basic
features are established in standard textbooks \cite{LanLif1991,Mer1998} for
particles confined by one-dimensional energy barriers: While the particle is
classically trapped, it can escape quantum mechanically if the energy barrier is
finite. This process typically exhibits an exponential decay $\exp{(-\gamma
t)}$, which is characterized by the tunneling rate $\gamma$. This rate describes
the inverse life-time of the confined state and captures the relevant
information of the tunneling process. Since time-independent one-dimensional
systems have integrable dynamics, the tunneling rates through the energy barrier
can be computed from complex WKB-paths in the classically forbidden region
\cite{LanLif1991,Mer1998,BerMou1972,Cre1994}
\begin{equation}
  \label{WKB}
  \gamma \propto \exp{\left(-\frac{2 \,\myIm{\,\Sif}}{\hbar} \right)}.
\end{equation}
Here, the imaginary part of the action $\Sif = \int p\,\text{d}q$, which
increases with the width and the height of the barrier, is divided by Planck's
constant $\heff = 2\pi\hbar$. In the semiclassical limit this ratio increases
and tunneling vanishes exponentially.

\begin{figure}[b]
  \begin{center}
    \includegraphics[width=\linewidth]{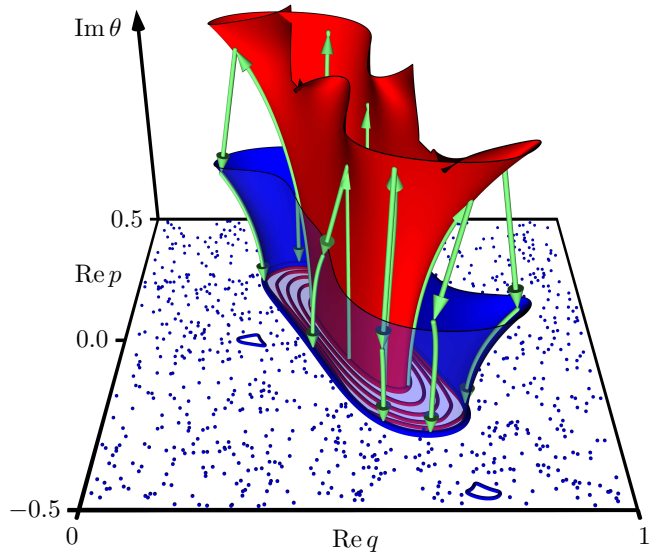}
    \caption{(color online) Dominant complex paths (green arrows) for the direct
regular-to-chaotic tunneling rate $\gamma_{1}$ of the standard map at $\kappa =
2.9$ and $\heff = 1/50$. The initial (inner red) torus and the final (outer
blue) torus of the fictitious integrable system emerge from the real classical
phase space.}
    \label{fig:complex_path_qp}
  \end{center}
\end{figure}

In contrast to one-dimensional potential wells, many systems of physical
relevance have non-integrable Hamiltonians such as driven atoms and molecules
\cite{ZacDelBuc1998, BucDelZac2002, WimSchEltBuc2006}, cold-atom systems
\cite{HenHafBroHecHelMcKMilPhiRolRubUpc2001, SteOskRai2001}, optical as well
as microwave cavities \cite{DemGraHeiHofRehRic2000,
BaeKetLoeRobVidHoeKuhSto2008, BaeKetLoeWieHen2009, Cre2007, CreWhi2012}, and
nano-wires \cite{FeiBaeKetRotHucBur2006}. These systems generically have a mixed
phase space in which classically disjoint regions of regular and chaotic motion
coexist. Quantum mechanical transitions between such regions are called
dynamical tunneling \cite{DavHel1981, KesSch2011}. In the above systems
dynamical tunneling is the key to understanding life-times and decay channels of
long-lived states associated to a regular region. Furthermore, dynamical
tunneling has important consequences for the structure of eigenfunctions
\cite{HufKetOttSch2002, BaeKetMon2005} and spectral statistics
\cite{VidStoRobKuhHoeGro2007, BatRob2010, BaeKetLoeMer2011, RudMerLoeBae2012} in
mixed systems. Due to this broad interest a lot of effort has been made to
investigate dynamical tunneling experimentally \cite{DemGraHeiHofRehRic2000,
BaeKetLoeRobVidHoeKuhSto2008, HenHafBroHecHelMcKMilPhiRolRubUpc2001,
SteOskRai2001} and theoretically \cite{HanOttAnt1984, Wil1986, BohTomUll1993,
TomUll1994, ShuIke1995, Cre1998, BroSchUll2001, PodNar2003, EltSch2005,
SheFisGuaReb2006, BaeKetLoeSch2008, Mou2007, ShuIshIke2002, LoeBaeKetSch2010,
DenMou2010}.

In this paper we derive a semiclassical prediction of dynamical tunneling rates
$\gamma$ for the ubiquitous situation of regular-to-chaotic tunneling. The
focus is on the experimentally relevant regime, in which direct tunneling to the
chaotic region dominates. We generalize the WKB-formula, Eq.~\eqref{WKB}, to
mixed systems by unifying the semiclassical time evolution method of complex
paths \cite{ShuIke1995} with the fictitious integrable system approach
\cite{BaeKetLoeSch2008}. This demonstrates that direct regular-to-chaotic
tunneling rates are determined by complex paths, which connect regular tori to
the boundary between the regular and the chaotic region, see
Fig.~\ref{fig:complex_path_qp}. This approach is successfully applied to the
standard map.

In the following we explain our semiclassical complex path approach to dynamical
tunneling rates for mixed systems. For simplicity we restrict the presentation
to time-periodic quantum systems, described by a time evolution operator~$\qmap$
over one period of the driving. The corresponding classical system is an
area-preserving map~$\map$.

\begin{figure}[tb]
  \begin{center}
    \includegraphics[]{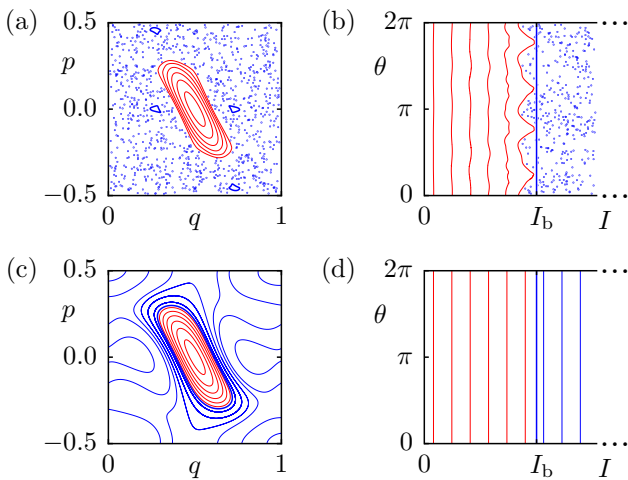}
    \caption{(color online) (a) Phase space of the standard map at $\kappa=2.9$
with regular tori (lines) and a chaotic orbit (dots), (b) in the action-angle
coordinates of $\Hreg$. (c) Phase space of $\Hreg$, (d) in action-angle
coordinates. The boundary torus $\Ib$ is marked by a thick line.}
    \label{fig:FISA}
  \end{center}
\end{figure}

We first introduce the fictitious integrable system approach
\cite{BaeKetLoeSch2008}. It is based on a fictitious integrable system
$\Hreg(q,p)$, which resembles the regular region of the mixed system as closely
as possible and extends it beyond its boundary, see Fig.~\ref{fig:FISA}(c). This
allows to decompose the Hilbert space into a regular and a chaotic part: As
basis states of these parts we use the eigenstates of $\Hreg$, which localize on
tori with quantized action $I_{\ireg} = \frac{1}{2 \pi}\oint p(q) \, \text{d}q =
\hbar(\ireg+\frac{1}{2})$. The action $\Ib$ of the boundary torus is used to
divide these states into regular basis states $|\Ii \rangle$ with $\Ii < \Ib$
and chaotic basis states with $\Ii \geq \Ib$, which will be called $|\If
\rangle$ in the following. Here, $\Ib$ is the (not necessarily quantized) action
of the first torus of $\Hreg$ that is entirely located in the chaotic region of
$\map$, see Fig.~\ref{fig:FISA}. In terms of the time evolution operator $\qmap$
of the mixed system, the rate $\gamma_{\ireg}$ of direct regular-to-chaotic
tunneling from the $\ireg$th quantizing torus to the chaotic region, is
determined by summing the transition probabilities to all chaotic basis states
\cite{BaeKetLoeSch2008}
\begin{equation}
  \label{FISA}
  \gamma_{\ireg} = \sum_{\ich} |\langle \If |\qmap |\Ii \rangle |^{2}.
\end{equation}

We now evaluate Eq.~\eqref{FISA} semiclassically, which will lead to our main
result Eq.~\eqref{gamma}. To this end we express the tunneling matrix elements
$\langle \If |\qmap |\Ii \rangle$ in the basis of the fictitious integrable
system by generalizing the time evolution technique of complex paths
\cite{ShuIke1995}. Introducing the position representation gives
\begin{equation}
  \label{FISA-projected}
  \langle \If| \qmap |\Ii\rangle = \int\text{d}\qf\int\text{d}\qi\hspace*{0.1cm}
  \langle \If|\qf \rangle \langle\qf |\qmap| \qi\rangle \langle\qi |\Ii\rangle.
\end{equation}
Here, the first and the last factor are the initial and the final wave function
of $\Hreg$. These two factors are replaced by their semiclassical counter-parts,
using the quantization of canonical transformations for the wave functions
\cite{Mil1974}. The propagator, $\langle\qf |\qmap| \qi\rangle$, is also
expressed semiclassically \cite{Sch2005, ShuIke1995}. The arising
integrals over $q$ and $q'$ are evaluated by the method of steepest descent
\cite{Mur1974}, i.e.,\ a saddle-point approximation, which explicitly allows to
take complex paths into account. This results in a semiclassical propagator for
the tunneling matrix elements
\begin{equation}
  \label{propagator}
  \begin{split}
 \langle \If| \qmap |\Ii\rangle = \hspace*{6.5cm}\\ \sum_{\paths}^{}
 \sqrt{\frac{\hbar}{2\pi}\frac{\partial^2 \Sifind{\paths} (\If,\Ii)}{\partial
\If \, \partial \Ii}} \myexp{\myi \left[ \frac{\Sifind{\paths}(\If,\Ii)}{\hbar}
+ \maslov_{\paths}\right]}.
\end{split}
\end{equation}
It is constructed from classical paths~$\paths$, with action $\Sifind{\paths}$
and a Maslov-phase shift $\maslov_{\paths}$. These paths~$\paths$ have to
connect the initial torus $\Ii$ to the final torus $\If$. Since there are no
such paths in real phase space, see Fig.~\ref{fig:FISA}, the above propagator is
exclusively constructed from paths of the complexified phase space. Such a
complex path $\paths$ consists of three segments: (i) The first segment is the
curve $\icurve_{\ireg,\paths}$ on the analytic continuation of the initial torus
$\Ii$ of $\Hreg$ into the complexified phase space. This curve connects the real
phase space to a specific point $(\qi_{\paths},\pin_{\paths})$ whose location
is determined by the next segment. (ii) The second segment is the classical path
of the complexified mixed system $\map$. It has to connect $(\qi_{\paths},
\pin_{\paths})$ on the initial complexified torus $\Ii$ to a point
$(\qf_{\paths}, \pf_{\paths})$ on the final complexified torus $\If$. This
requirement determines the point $(\qi_{\paths}, \pin_{\paths})$ of the first
segment. (iii) The final segment is the curve $\fcurve_{\ich, \paths}$ which
connects the point $(\qf_{\paths}, \pf_{\paths})$ back to the real phase space,
along the complexified final torus $\If$ of $\Hreg$. The three segments are
sketched in Fig.~\ref{fig:complex_path_qp} (green curves). Their explicit
determination for the standard map will be described below.

The action $\Sifind{\paths}$ of such complex paths is given by
\begin{equation}
\label{action}
\Sifind{\paths}(\If,\Ii) = \hspace*{-0.25cm} \Si \,+
\,\Smapind{\paths}(\qi_{\paths},\qf_{\paths})\, + \hspace*{-0.2cm}
\Sf.
\end{equation}
Here, the first and the last contribution are action integrals carried out along
the curves $\icurve_{\ireg,\paths}$ and $\fcurve_{\ich,\paths}$, respectively.
The action $\Smapind{\paths}(\qi_{\paths},\qf_{\paths})$ originates from the
second segment, where the mixed system propagates between the initial and the
final tori. From the possibly infinite number of paths, which contribute to the
sum in Eq.~\eqref{propagator}, we select the dominant ones having the smallest
positive imaginary action. To obtain the relative phase between the paths
$\paths$, the curves $\icurve_{\ireg,\paths}$ have to start at common reference
points and the curves $\fcurve_{\ich,\paths}$ have to end at common reference
points in real phase space.

In the following we evaluate the tunneling rates in Eq.~\eqref{FISA}
semiclassically. To this end, we replace the sum $\sum_{\text{\ich}}$ by an
integral $\int_{\Ib}^{\infty} \text{d}\If/\hbar$, starting from the boundary
action $\Ib$. Here, we exploit that transition matrix elements are
semiclassically not restricted to quantizing actions. The matrix elements,
Eq.~\eqref{propagator}, are maximal at the boundary $\Ib$ and decrease
exponentially with increasing $\If$. This allows to approximate the integral
asymptotically \cite{Mur1974}, leading to our final semiclassical result for
dynamical tunneling rates
\begin{equation}
  \label{gamma}
  \gamma_{m} = \sum\limits_{\paths}^{}
 	  \frac{\hbar}{4\pi} \frac{ \left| \frac{\partial^2 \Sif_{\paths}
 (\Ib, \Ii) }{\partial \Ib \partial \Ii} \right|}{\myIm{ \frac{\partial
 \Sif_{\paths} (\Ib,\Ii)}{\partial \Ib } }}
 \myexp{-
\frac{2\,\myIm{\,\Sif_{\paths} (\Ib, \Ii)}}{\hbar} }.
\end{equation}
Here, the only relevant paths are those which connect the regular action $\Ii$
to the boundary action $\Ib$. Moreover, it generalizes the familiar
WKB-prediction of Eq.~\eqref{WKB}. The semiclassical approximation leading to
Eq.~\eqref{gamma} cancels the interference terms between different paths. Hence,
it is not necessary to determine the relative phases of the individual paths for
a semiclassical prediction of dynamical tunneling rates.

\begin{figure}[tb]
  \begin{center}
    \includegraphics[width=\linewidth]{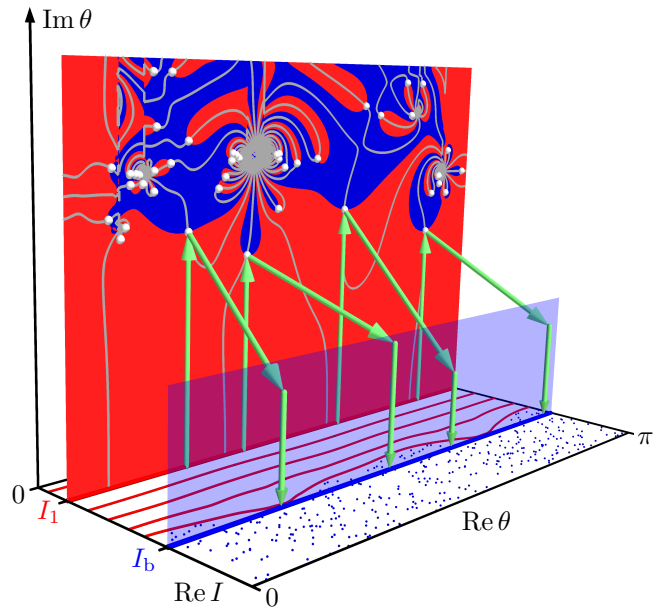}
    \caption{(color online) Dominant complex paths (green arrows) for the
tunneling rate $\gamma_{1}$ of the standard map at $\kappa = 2.9$ and $\heff =
1/50$ as in Fig.~\ref{fig:complex_path_qp} but in action-angle representation.
The planes are the complexifications of the initial torus $I_{1}$ and the final
torus $\Ib$ of $\Hreg$. White dots represent points on $I_{1}$ which map onto
$\Ib$. They are at the intersection of the gray lines ($\myIm{\,\Ef} = 0$) with
the boundary between light red regions ($\myRe{\,\Ef} < \Eb$) and dark blue
regions ($\myRe{\,\Ef} > \Eb$). The visualization is simplified by shifting the
angle of the endpoints of step (ii) by $-\omega_{\text{b}}$ along $\Ib$ [as
also done in Fig.~\ref{fig:complex_path_qp}].
}
    \label{fig:complex_path_actang}
  \end{center}
\end{figure}

We illustrate our approach and demonstrate its successful predictions by
applying it to the standard map \cite{Chi1979}, which is the paradigmatic
example of a mixed system. It is obtained from the one-dimensional kicked
Hamiltonian $H(q,p) = T(p)+V(q)\sum_{n}\delta{(t-n)}$ with $T(p) = p^2/2$
and $V(q) = \kappa/(2\pi)^2\cos{(2\pi q)}$. The stroboscopic time evolution is
given by the map $\map$
\begin{align}
  \label{standardmap}
 \qf = \qi + T'(\pin) \;\;\;\;\;\;\;\; \pf = \pin -V'(\qf).
\end{align}
The quantum time-evolution operator $\qmap$ is given by $\qmap = \myexp{-\myi
V(q)/\hbar}\myexp{-\myi T(p)/\hbar}$, where $\heff = 2\pi\hbar$ is the
effective Planck constant. We determine the tunneling rate $\gamma_{\ireg}$ of
the $\ireg$th quantizing torus semiclassically, using Eq.~\eqref{gamma}: One
constructs a fictitious integrable system $\Hreg(q,p)$, Fig.~\ref{fig:FISA}(c),
based on the frequencies of the main regular island \cite{BaeKetLoeSch2008}. For
this $\Hreg(q,p)$ there is a canonical transformation to action-angle
coordinates $(\act,\ang)$, see Fig.~\ref{fig:FISA}(d), giving $\Hreg(\act)$.
This allows for searching the complex paths $\paths$ which connect the real
initial torus $\Ii$ and the real boundary torus $\Ib$ of $\Hreg$, according to
steps (i)-(iii).
For step (i) the torus $\Ii$ is analytically continued by
complexifying the angle $\ang$. This torus forms a plane in the complexified
phase space, see Fig.~\ref{fig:complex_path_actang}, parametrized by the real
and the imaginary part of the angle $\ang$. To obtain the canonical
transformation to the corresponding $(q,p)$-coordinates, we numerically
integrate Hamiltons equations for $\Hreg(q,p)$ in imaginary time, starting
from the real torus $\Ii$. The integration is done up to time
$t=\text{i}\,\myIm{(\ang)}/\omega_{\ireg}$ where $\omega_{\ireg}$ is the
frequency of the real torus $\Ii$.
For step (ii) the points $(q,p)$ of the complexified initial torus $\Ii$ are
mapped by the complexification of the standard map, Eq.~\eqref{standardmap}, to
points $(\qf,\pf)$. The points $(q,p)$ are mapped onto the complexified torus
$\Ib$, only if the energy $\Ef = \Hreg(\qf,\pf)$ equals the energy
$\Eb$ of the boundary torus $\Ib$. These numerically determined points are
labeled $(\qi_{\paths},\pin_{\paths})$ and are shown in
Fig.~\ref{fig:complex_path_actang} as white dots.
For step (iii) we integrate Hamiltons equations for $\Hreg$ in negative
imaginary time to get back to the real torus $\Ib$. Combining the three steps
(i)-(iii) gives the complex paths $\paths$.
Finally we evaluate the action of these paths $\paths$ according to
Eq.~\eqref{action} in which the second contribution for the standard map is
given by \cite{BerBalTabVor1979, Bog1992}
\begin{equation}
  \Smapind{}(\qi_{\paths},\qf_{\paths}) =
\frac{(\qi_{\paths}-\qf_{\paths})^2}{2} - V(\qf_{\paths}).
\end{equation}
From the infinite number of paths~$\paths$ contributing to the sum in
Eq.~\eqref{gamma} we select the dominant ones, which have the smallest positive
imaginary action, see arrows in Fig.~\ref{fig:complex_path_actang}. There, the
four paths~$\paths$
(together with their symmetry partners) contribute 20~\%, 6~\%, 33~\%, and 41~\%
(left to right) to the tunneling rate $\gamma_{1}$. Note, that we account for
the parity of the standard map by just considering one symmetry partner and
doubling its contribution in Eq.~\eqref{propagator}, which leads to an
additional factor of four in Eq.~\eqref{gamma}.

\begin{figure}[!tb]
  \begin{center}
    \includegraphics[]{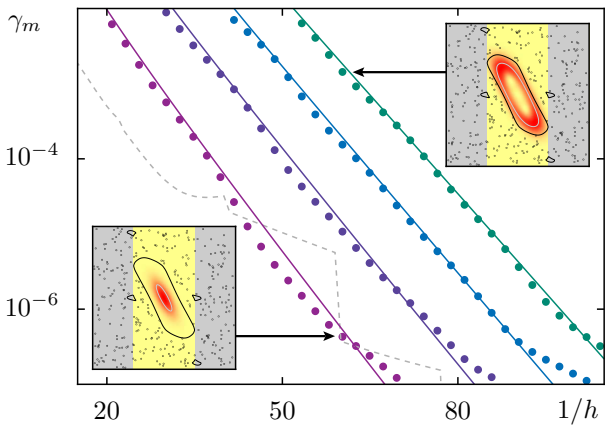}
    \caption{(color online) Tunneling rates $\gamma_{m}$ of the standard
map at $\kappa=2.9$ for $\ireg$ = $0,1,2,3$ vs.\ $1/\heff$. The semiclassical
prediction, Eq.~\eqref{gamma}, (straight lines) is compared to numerical rates
(dots) and for $m=0$ to the quantum evaluation of Eq.~\eqref{FISA} (dashed
line). The insets show the Husimi representation of the eigenstates with $\ireg$
= $0$ and $\ireg$ = $3$ at $\heff$ = $1/60$ and the corresponding quantizing
tori (gray lines) in phase space with absorbing regions (gray).}
    \label{fig:results}
  \end{center}
\end{figure}
The semiclassically predicted tunneling rates $\gamma_{\ireg}$,
Eq.~\eqref{gamma}, for the standard map are shown as lines in
Fig.~\ref{fig:results}. Numerically determined tunneling rates (dots) are
obtained by opening the system \cite{BaeKetLoeSch2008} tangential to the regular
region (inset in Fig.~\ref{fig:results}), which is consistent with the choice of
$\Ib$ (Fig.~\ref{fig:FISA}). The semiclassical prediction accurately describes
the numerical tunneling rates, with deviations by less than a factor of 2. This
considerably improves the quality of previous predictions
\cite{BaeKetLoeSch2008}, in which one finds quantization steps from the quantum
mechanical evaluation of Eq.~\eqref{FISA}, see Fig.~\ref{fig:results}. These
quantization steps do not occur in the semiclassical evaluation, as $\Ib$ does
not have to obey any quantization condition. The small deviations at small
$\gamma$ can be attributed to non-linear resonances. One should be able take
this effect into account, by taking our semiclassical prediction for direct
regular-to-chaotic tunneling and combine \cite{LoeBaeKetSch2010} it with
resonance-assisted tunneling.

We now make a couple of remarks:
(a) Our complex-path approach also works for other parameters $\kappa$ of the
standard map, leading to predictions of similar quality. Deviations due to
nonlinear resonances may arise at larger $\gamma$ already.
(b) Partial barriers which limit transport in the chaotic region can
additionally suppress dynamical tunneling \cite{BohTomUll1993}. This
effect is not yet considered in our investigations since we open the system at
the boundary of the regular region.
(c) We obtain the tunneling rates $\gamma_{m}$ from semiclassical propagation
over a single time step and find that only a small number of complex
paths $\paths$ contribute to Eq.~\eqref{gamma}. Thus the determination of
tunneling rates turns out to be considerably simpler than the semiclassical
propagation of wave packets over long times, for which a careful selection among
hundreds of dominant paths is needed \cite{ShuIke1995}.
(d) The natural boundary \cite{GrePer1981, Per1982} makes it impossible
to continue regular tori of non-integrable systems deep into the complexified
phase space \cite{Cre1998, Cre2007, CreWhi2012}. We overcome this problem by
complexifying the tori of the fictitious integrable system $\Hreg$.
(e) The choice of $\Hreg$ is not unique and defining a precise classical
criterion for its quality is an open question. While the paths~$\paths$ may
depend on this choice, we expect to obtain a prediction of tunneling rates
$\gamma_{m}$ with similar agreement to numerical rates.
(f) Determining tunneling rates just from WKB-paths along the initial torus
$\Ii$ of $\Hreg$, which was successful in specific situations
\cite{BroSchUll2001, SheFisGuaReb2006, Mou2007, BaeKetLoeSch2008}, gives for the
standard map a rough estimate only.

In summary, we have presented a complex-path approach which allows for
predicting regular-to-chaotic tunneling rates in systems with a mixed phase
space. We have successfully applied this method to predict tunneling rates of
the standard map.
For the future it is desirable to develop a complete semiclassical description
of regular-to-chaotic tunneling even in the presence of resonance-assisted
tunneling.
Moreover, the application of our approach to higher dimensional systems like
billiards, optical microcavities, or atoms and molecules remains an open
problem.
Finally, we believe that our approach is the basis to reveal the universal
classical properties which govern dynamical tunneling.

We acknowledge helpful discussions with S.~Creagh and K.~Ikeda and the
participants of the Advanced Study Group "Towards a Semiclassical Theory of
Dynamical Tunneling". We thank the DFG for support within Forschergruppe 760
"Scattering Systems with Complex Dynamics". AB thanks the Japan Society for the
Promotion of Science for supporting his stay in Japan.


\begin{thebibliography}{54}
\expandafter\ifx\csname natexlab\endcsname\relax\def\natexlab#1{#1}\fi
\expandafter\ifx\csname bibnamefont\endcsname\relax
  \def\bibnamefont#1{#1}\fi
\expandafter\ifx\csname bibfnamefont\endcsname\relax
  \def\bibfnamefont#1{#1}\fi
\expandafter\ifx\csname citenamefont\endcsname\relax
  \def\citenamefont#1{#1}\fi
\expandafter\ifx\csname url\endcsname\relax
  \def\url#1{\texttt{#1}}\fi
\expandafter\ifx\csname urlprefix\endcsname\relax\def\urlprefix{URL }\fi
\providecommand{\bibinfo}[2]{#2}
\providecommand{\eprint}[2][]{\url{#2}}

\bibitem[{\citenamefont{Landau and Lifshitz}(1991)}]{LanLif1991}
\bibinfo{author}{\bibfnamefont{L.~D.} \bibnamefont{Landau}} \bibnamefont{and}
  \bibinfo{author}{\bibfnamefont{E.~M.} \bibnamefont{Lifshitz}},
  \emph{\bibinfo{title}{Course of Theoretical Physics, Vol.3: Quantum
  Mechanics}} (\bibinfo{publisher}{Pergamon Press, New York},
  \bibinfo{year}{1991}).

\bibitem[{\citenamefont{Merzbacher}(1998)}]{Mer1998}
\bibinfo{author}{\bibfnamefont{E.}~\bibnamefont{Merzbacher}},
  \emph{\bibinfo{title}{Quantum mechanics}} (\bibinfo{publisher}{Wiley, New
  York}, \bibinfo{year}{1998}).

\bibitem[{\citenamefont{Berry and Mount}(1972)}]{BerMou1972}
\bibinfo{author}{\bibfnamefont{M.~V.} \bibnamefont{Berry}} \bibnamefont{and}
  \bibinfo{author}{\bibfnamefont{K.~E.} \bibnamefont{Mount}},
  \bibinfo{journal}{Rep. Proc. Phys.} \textbf{\bibinfo{volume}{35}},
  \bibinfo{pages}{315} (\bibinfo{year}{1972}).

\bibitem[{\citenamefont{Creagh}(1994)}]{Cre1994}
\bibinfo{author}{\bibfnamefont{S.~C.} \bibnamefont{Creagh}},
  \bibinfo{journal}{J. Phys. A.} \textbf{\bibinfo{volume}{27}},
  \bibinfo{pages}{4969} (\bibinfo{year}{1994}).

\bibitem[{\citenamefont{Zakrzewski et~al.}(1998)\citenamefont{Zakrzewski,
  Delande, and Buchleitner}}]{ZacDelBuc1998}
\bibinfo{author}{\bibfnamefont{J.}~\bibnamefont{Zakrzewski}},
  \bibinfo{author}{\bibfnamefont{D.}~\bibnamefont{Delande}}, \bibnamefont{and}
  \bibinfo{author}{\bibfnamefont{A.}~\bibnamefont{Buchleitner}},
  \bibinfo{journal}{Phys. Rev. E} \textbf{\bibinfo{volume}{57}},
  \bibinfo{pages}{1458} (\bibinfo{year}{1998}).

\bibitem[{\citenamefont{Buchleitner et~al.}(2002)\citenamefont{Buchleitner,
  Delande, and Zakrzewski}}]{BucDelZac2002}
\bibinfo{author}{\bibfnamefont{A.}~\bibnamefont{Buchleitner}},
  \bibinfo{author}{\bibfnamefont{D.}~\bibnamefont{Delande}}, \bibnamefont{and}
  \bibinfo{author}{\bibfnamefont{J.}~\bibnamefont{Zakrzewski}},
  \bibinfo{journal}{Phys. Rep.} \textbf{\bibinfo{volume}{368}},
  \bibinfo{pages}{409} (\bibinfo{year}{2002}).

\bibitem[{\citenamefont{Wimberger et~al.}(2006)\citenamefont{Wimberger,
  Schlagheck, Eltschka, and Buchleitner}}]{WimSchEltBuc2006}
\bibinfo{author}{\bibfnamefont{S.}~\bibnamefont{Wimberger}},
  \bibinfo{author}{\bibfnamefont{P.}~\bibnamefont{Schlagheck}},
  \bibinfo{author}{\bibfnamefont{C.}~\bibnamefont{Eltschka}}, \bibnamefont{and}
  \bibinfo{author}{\bibfnamefont{A.}~\bibnamefont{Buchleitner}},
  \bibinfo{journal}{Phys. Rev. Lett.} \textbf{\bibinfo{volume}{97}},
  \bibinfo{eid}{043001} (\bibinfo{year}{2006}).

\bibitem[{\citenamefont{Hensinger et~al.}(2001)\citenamefont{Hensinger,
  H\"affner, Browaeys, Heckenberg, Helmerson, McKenzie, Milburn, Phillips,
  Rolston, Rubinsztein-Dunlop et~al.}}]{HenHafBroHecHelMcKMilPhiRolRubUpc2001}
\bibinfo{author}{\bibfnamefont{W.~K.} \bibnamefont{Hensinger}},
  \bibinfo{author}{\bibfnamefont{H.}~\bibnamefont{H\"affner}},
  \bibinfo{author}{\bibfnamefont{A.}~\bibnamefont{Browaeys}},
  \bibinfo{author}{\bibfnamefont{N.~R.} \bibnamefont{Heckenberg}},
  \bibinfo{author}{\bibfnamefont{K.}~\bibnamefont{Helmerson}},
  \bibinfo{author}{\bibfnamefont{C.}~\bibnamefont{McKenzie}},
  \bibinfo{author}{\bibfnamefont{G.~J.} \bibnamefont{Milburn}},
  \bibinfo{author}{\bibfnamefont{W.~D.} \bibnamefont{Phillips}},
  \bibinfo{author}{\bibfnamefont{S.~L.} \bibnamefont{Rolston}},
  \bibinfo{author}{\bibfnamefont{H.}~\bibnamefont{Rubinsztein-Dunlop}},
  \bibnamefont{et~al.}, \bibinfo{journal}{Nature}
  \textbf{\bibinfo{volume}{412}}, \bibinfo{pages}{52} (\bibinfo{year}{2001}).

\bibitem[{\citenamefont{Steck et~al.}(2001)\citenamefont{Steck, Oskay, and
  Raizen}}]{SteOskRai2001}
\bibinfo{author}{\bibfnamefont{D.~A.} \bibnamefont{Steck}},
  \bibinfo{author}{\bibfnamefont{W.~H.} \bibnamefont{Oskay}}, \bibnamefont{and}
  \bibinfo{author}{\bibfnamefont{M.~G.} \bibnamefont{Raizen}},
  \bibinfo{journal}{Science} \textbf{\bibinfo{volume}{293}},
  \bibinfo{pages}{274} (\bibinfo{year}{2001}).

\bibitem[{\citenamefont{Dembowski et~al.}(2000)\citenamefont{Dembowski, Gr\"af,
  Heine, Hofferbert, Rehfeld, and Richter}}]{DemGraHeiHofRehRic2000}
\bibinfo{author}{\bibfnamefont{C.}~\bibnamefont{Dembowski}},
  \bibinfo{author}{\bibfnamefont{H.-D.} \bibnamefont{Gr\"af}},
  \bibinfo{author}{\bibfnamefont{A.}~\bibnamefont{Heine}},
  \bibinfo{author}{\bibfnamefont{R.}~\bibnamefont{Hofferbert}},
  \bibinfo{author}{\bibfnamefont{H.}~\bibnamefont{Rehfeld}}, \bibnamefont{and}
  \bibinfo{author}{\bibfnamefont{A.}~\bibnamefont{Richter}},
  \bibinfo{journal}{Phys. Rev. Lett.} \textbf{\bibinfo{volume}{84}},
  \bibinfo{pages}{867} (\bibinfo{year}{2000}).

\bibitem[{\citenamefont{B\"acker
  et~al.}(2008{\natexlab{a}})\citenamefont{B\"acker, Ketzmerick, L\"ock,
  Robnik, Vidmar, H\"ohmann, Kuhl, and
  St\"ockmann}}]{BaeKetLoeRobVidHoeKuhSto2008}
\bibinfo{author}{\bibfnamefont{A.}~\bibnamefont{B\"acker}},
  \bibinfo{author}{\bibfnamefont{R.}~\bibnamefont{Ketzmerick}},
  \bibinfo{author}{\bibfnamefont{S.}~\bibnamefont{L\"ock}},
  \bibinfo{author}{\bibfnamefont{M.}~\bibnamefont{Robnik}},
  \bibinfo{author}{\bibfnamefont{G.}~\bibnamefont{Vidmar}},
  \bibinfo{author}{\bibfnamefont{R.}~\bibnamefont{H\"ohmann}},
  \bibinfo{author}{\bibfnamefont{U.}~\bibnamefont{Kuhl}}, \bibnamefont{and}
  \bibinfo{author}{\bibfnamefont{H.-J.} \bibnamefont{St\"ockmann}},
  \bibinfo{journal}{Phys. Rev. Lett.} \textbf{\bibinfo{volume}{100}},
  \bibinfo{pages}{174103} (\bibinfo{year}{2008}{\natexlab{a}}).

\bibitem[{\citenamefont{Creagh}(2007)}]{Cre2007}
\bibinfo{author}{\bibfnamefont{S.~C.} \bibnamefont{Creagh}},
  \bibinfo{journal}{Phys. Rev. Lett.} \textbf{\bibinfo{volume}{98}},
  \bibinfo{pages}{153901} (\bibinfo{year}{2007}).

\bibitem[{\citenamefont{B\"acker et~al.}(2009)\citenamefont{B\"acker,
  Ketzmerick, L\"ock, Wiersig, and Hentschel}}]{BaeKetLoeWieHen2009}
\bibinfo{author}{\bibfnamefont{A.}~\bibnamefont{B\"acker}},
  \bibinfo{author}{\bibfnamefont{R.}~\bibnamefont{Ketzmerick}},
  \bibinfo{author}{\bibfnamefont{S.}~\bibnamefont{L\"ock}},
  \bibinfo{author}{\bibfnamefont{J.}~\bibnamefont{Wiersig}}, \bibnamefont{and}
  \bibinfo{author}{\bibfnamefont{M.}~\bibnamefont{Hentschel}},
  \bibinfo{journal}{Phys. Rev. A} \textbf{\bibinfo{volume}{79}},
  \bibinfo{pages}{063804} (\bibinfo{year}{2009}).  
  
\bibitem[{\citenamefont{Creagh and White}(2012)}]{CreWhi2012}
\bibinfo{author}{\bibfnamefont{S.~C.} \bibnamefont{Creagh}} \bibnamefont{and}
  \bibinfo{author}{\bibfnamefont{M.~M.} \bibnamefont{White}},
  \bibinfo{journal}{Phys. Rev. E} \textbf{\bibinfo{volume}{85}},
  \bibinfo{pages}{015201} (\bibinfo{year}{2012}).

\bibitem[{\citenamefont{Feist et~al.}(2006)\citenamefont{Feist, B\"acker,
  Ketzmerick, Rotter, Huckestein, and Burgd\"orfer}}]{FeiBaeKetRotHucBur2006}
\bibinfo{author}{\bibfnamefont{J.}~\bibnamefont{Feist}},
  \bibinfo{author}{\bibfnamefont{A.}~\bibnamefont{B\"acker}},
  \bibinfo{author}{\bibfnamefont{R.}~\bibnamefont{Ketzmerick}},
  \bibinfo{author}{\bibfnamefont{S.}~\bibnamefont{Rotter}},
  \bibinfo{author}{\bibfnamefont{B.}~\bibnamefont{Huckestein}},
  \bibnamefont{and}
  \bibinfo{author}{\bibfnamefont{J.}~\bibnamefont{Burgd\"orfer}},
  \bibinfo{journal}{Phys. Rev. Lett.} \textbf{\bibinfo{volume}{97}},
  \bibinfo{pages}{116804} (\bibinfo{year}{2006});
%
\bibinfo{author}{\bibfnamefont{J.}~\bibnamefont{Feist}},
  \bibinfo{author}{\bibfnamefont{A.}~\bibnamefont{B\"acker}},
  \bibinfo{author}{\bibfnamefont{R.}~\bibnamefont{Ketzmerick}},
  \bibinfo{author}{\bibfnamefont{J.}~\bibnamefont{Burgd\"orfer}},
  \bibnamefont{and} \bibinfo{author}{\bibfnamefont{S.}~\bibnamefont{Rotter}},
  \bibinfo{journal}{Phys. Rev. B} \textbf{\bibinfo{volume}{80}},
  \bibinfo{pages}{245322} (\bibinfo{year}{2009}).

\bibitem[{\citenamefont{Davis and Heller}(1981)}]{DavHel1981}
\bibinfo{author}{\bibfnamefont{M.~J.} \bibnamefont{Davis}} \bibnamefont{and}
  \bibinfo{author}{\bibfnamefont{E.~J.} \bibnamefont{Heller}},
  \bibinfo{journal}{J. Chem. Phys.} \textbf{\bibinfo{volume}{75}},
  \bibinfo{pages}{246} (\bibinfo{year}{1981}).

\bibitem[{\citenamefont{Keshavamurthy and Schlagheck}(2011)}]{KesSch2011}
\bibinfo{author}{\bibfnamefont{S.}~\bibnamefont{Keshavamurthy}}
  \bibnamefont{and}
  \bibinfo{author}{\bibfnamefont{P.}~\bibnamefont{Schlagheck}},
  \emph{\bibinfo{title}{Dynamical Tunneling: Theory and Experiment}}
  (\bibinfo{publisher}{CRC Press, New York}, \bibinfo{year}{2011}).

\bibitem[{\citenamefont{Hufnagel et~al.}(2002)\citenamefont{Hufnagel,
  Ketzmerick, Otto, and Schanz}}]{HufKetOttSch2002}
\bibinfo{author}{\bibfnamefont{L.}~\bibnamefont{Hufnagel}},
  \bibinfo{author}{\bibfnamefont{R.}~\bibnamefont{Ketzmerick}},
  \bibinfo{author}{\bibfnamefont{M.-F.} \bibnamefont{Otto}}, \bibnamefont{and}
  \bibinfo{author}{\bibfnamefont{H.}~\bibnamefont{Schanz}},
  \bibinfo{journal}{Phys. Rev. Lett.} \textbf{\bibinfo{volume}{89}},
  \bibinfo{pages}{154101} (\bibinfo{year}{2002}).

\bibitem[{\citenamefont{B\"acker et~al.}(2005)\citenamefont{B\"acker,
  Ketzmerick, and Monastra}}]{BaeKetMon2005}
\bibinfo{author}{\bibfnamefont{A.}~\bibnamefont{B\"acker}},
  \bibinfo{author}{\bibfnamefont{R.}~\bibnamefont{Ketzmerick}},
  \bibnamefont{and} \bibinfo{author}{\bibfnamefont{A.~G.}
  \bibnamefont{Monastra}}, \bibinfo{journal}{Phys. Rev. Lett.}
  \textbf{\bibinfo{volume}{94}}, \bibinfo{pages}{054102}
  (\bibinfo{year}{2005});
%
\bibinfo{journal}{Phys. Rev. E}
  \textbf{\bibinfo{volume}{75}}, \bibinfo{pages}{066204}
  (\bibinfo{year}{2007}).

\bibitem[{\citenamefont{Vidmar et~al.}(2007)\citenamefont{Vidmar,
  St\"{o}ckmann, Robnik, Kuhl, H\"{o}hmann, and
  Grossmann}}]{VidStoRobKuhHoeGro2007}
\bibinfo{author}{\bibfnamefont{G.}~\bibnamefont{Vidmar}},
  \bibinfo{author}{\bibfnamefont{H.-J.} \bibnamefont{St\"{o}ckmann}},
  \bibinfo{author}{\bibfnamefont{M.}~\bibnamefont{Robnik}},
  \bibinfo{author}{\bibfnamefont{U.}~\bibnamefont{Kuhl}},
  \bibinfo{author}{\bibfnamefont{R.}~\bibnamefont{H\"{o}hmann}},
  \bibnamefont{and}
  \bibinfo{author}{\bibfnamefont{S.}~\bibnamefont{Grossmann}},
  \bibinfo{journal}{J. Phys. A} \textbf{\bibinfo{volume}{40}},
  \bibinfo{pages}{13883} (\bibinfo{year}{2007}).

\bibitem[{\citenamefont{Batisti{\'c} and Robnik}(2010)}]{BatRob2010}
\bibinfo{author}{\bibfnamefont{B.}~\bibnamefont{Batisti{\'c}}}
  \bibnamefont{and} \bibinfo{author}{\bibfnamefont{M.}~\bibnamefont{Robnik}},
  \bibinfo{journal}{J. Phys. A} \textbf{\bibinfo{volume}{43}},
  \bibinfo{pages}{215101} (\bibinfo{year}{2010}).

\bibitem[{\citenamefont{B\"acker et~al.}(2011)\citenamefont{B\"acker,
  Ketzmerick, L\"ock, and Mertig}}]{BaeKetLoeMer2011}
\bibinfo{author}{\bibfnamefont{A.}~\bibnamefont{B\"acker}},
  \bibinfo{author}{\bibfnamefont{R.}~\bibnamefont{Ketzmerick}},
  \bibinfo{author}{\bibfnamefont{S.}~\bibnamefont{L\"ock}}, \bibnamefont{and}
  \bibinfo{author}{\bibfnamefont{N.}~\bibnamefont{Mertig}},
  \bibinfo{journal}{Phys. Rev. Lett.} \textbf{\bibinfo{volume}{106}},
  \bibinfo{pages}{024101} (\bibinfo{year}{2011}).

\bibitem[{\citenamefont{Rudolf et~al.}(2012)\citenamefont{Rudolf, Mertig,
  L\"ock, and B\"acker}}]{RudMerLoeBae2012}
\bibinfo{author}{\bibfnamefont{T.}~\bibnamefont{Rudolf}},
  \bibinfo{author}{\bibfnamefont{N.}~\bibnamefont{Mertig}},
  \bibinfo{author}{\bibfnamefont{S.}~\bibnamefont{L\"ock}}, \bibnamefont{and}
  \bibinfo{author}{\bibfnamefont{A.}~\bibnamefont{B\"acker}},
  \bibinfo{journal}{Phys. Rev. E} \textbf{\bibinfo{volume}{85}},
  \bibinfo{pages}{036213} (\bibinfo{year}{2012}).

\bibitem[{\citenamefont{Hanson et~al.}(1984)\citenamefont{Hanson, Ott, and
  Antonsen}}]{HanOttAnt1984}
\bibinfo{author}{\bibfnamefont{J.~D.} \bibnamefont{Hanson}},
  \bibinfo{author}{\bibfnamefont{E.}~\bibnamefont{Ott}}, \bibnamefont{and}
  \bibinfo{author}{\bibfnamefont{T.~M.} \bibnamefont{Antonsen}},
  \bibinfo{journal}{Phys. Rev. A} \textbf{\bibinfo{volume}{29}},
  \bibinfo{pages}{819} (\bibinfo{year}{1984}).

\bibitem[{\citenamefont{Wilkinson}(1986)}]{Wil1986}
\bibinfo{author}{\bibfnamefont{M.}~\bibnamefont{Wilkinson}},
  \bibinfo{journal}{Physica D} \textbf{\bibinfo{volume}{21}},
  \bibinfo{pages}{341} (\bibinfo{year}{1986}).

\bibitem[{\citenamefont{Bohigas et~al.}(1993)\citenamefont{Bohigas, Tomsovic,
  and Ullmo}}]{BohTomUll1993}
\bibinfo{author}{\bibfnamefont{O.}~\bibnamefont{Bohigas}},
  \bibinfo{author}{\bibfnamefont{S.}~\bibnamefont{Tomsovic}}, \bibnamefont{and}
  \bibinfo{author}{\bibfnamefont{D.}~\bibnamefont{Ullmo}},
  \bibinfo{journal}{Phys. Rep.} \textbf{\bibinfo{volume}{223}},
  \bibinfo{pages}{43} (\bibinfo{year}{1993}).

\bibitem[{\citenamefont{Tomsovic and Ullmo}(1994)}]{TomUll1994}
\bibinfo{author}{\bibfnamefont{S.}~\bibnamefont{Tomsovic}} \bibnamefont{and}
  \bibinfo{author}{\bibfnamefont{D.}~\bibnamefont{Ullmo}},
  \bibinfo{journal}{Phys. Rev. E} \textbf{\bibinfo{volume}{50}},
  \bibinfo{pages}{145} (\bibinfo{year}{1994}).

\bibitem[{\citenamefont{Shudo and Ikeda}(1995)}]{ShuIke1995}
\bibinfo{author}{\bibfnamefont{A.}~\bibnamefont{Shudo}} \bibnamefont{and}
  \bibinfo{author}{\bibfnamefont{K.~S.} \bibnamefont{Ikeda}},
  \bibinfo{journal}{Phys. Rev. Lett.} \textbf{\bibinfo{volume}{74}},
  \bibinfo{pages}{682} (\bibinfo{year}{1995});
%
  \bibinfo{journal}{Physica D} \textbf{\bibinfo{volume}{115}},
  \bibinfo{pages}{234} (\bibinfo{year}{1998}).

\bibitem[{\citenamefont{Creagh}(1998)}]{Cre1998}
\bibinfo{author}{\bibfnamefont{S.~C.} \bibnamefont{Creagh}},
  \emph{\bibinfo{title}{Tunneling in two dimensions}}, in {\it Tunneling in
  Complex Systems} (\bibinfo{publisher}{World Scientific, Singapore},
  \bibinfo{year}{1998}).

\bibitem[{\citenamefont{Brodier et~al.}(2001)\citenamefont{Brodier, Schlagheck,
  and Ullmo}}]{BroSchUll2001}
\bibinfo{author}{\bibfnamefont{O.}~\bibnamefont{Brodier}},
  \bibinfo{author}{\bibfnamefont{P.}~\bibnamefont{Schlagheck}},
  \bibnamefont{and} \bibinfo{author}{\bibfnamefont{D.}~\bibnamefont{Ullmo}},
  \bibinfo{journal}{Phys. Rev. Lett.} \textbf{\bibinfo{volume}{87}},
  \bibinfo{pages}{064101} (\bibinfo{year}{2001});
%
  \bibinfo{journal}{Ann. of Phys.} \textbf{\bibinfo{volume}{300}},
  \bibinfo{pages}{88} (\bibinfo{year}{2002}).

\bibitem[{\citenamefont{Podolskiy and Narimanov}(2003)}]{PodNar2003}
\bibinfo{author}{\bibfnamefont{V.~A.} \bibnamefont{Podolskiy}}
  \bibnamefont{and} \bibinfo{author}{\bibfnamefont{E.~E.}
  \bibnamefont{Narimanov}}, \bibinfo{journal}{Phys. Rev. Lett.}
  \textbf{\bibinfo{volume}{91}}, \bibinfo{pages}{263601}
  (\bibinfo{year}{2003}).

\bibitem[{\citenamefont{Eltschka and Schlagheck}(2005)}]{EltSch2005}
\bibinfo{author}{\bibfnamefont{C.}~\bibnamefont{Eltschka}} \bibnamefont{and}
  \bibinfo{author}{\bibfnamefont{P.}~\bibnamefont{Schlagheck}},
  \bibinfo{journal}{Phys. Rev. Lett.} \textbf{\bibinfo{volume}{94}},
  \bibinfo{pages}{014101} (\bibinfo{year}{2005}).

\bibitem[{\citenamefont{Sheinman et~al.}(2006)\citenamefont{Sheinman, Fishman,
  Guarneri, and Rebuzzini}}]{SheFisGuaReb2006}
\bibinfo{author}{\bibfnamefont{M.}~\bibnamefont{Sheinman}},
  \bibinfo{author}{\bibfnamefont{S.}~\bibnamefont{Fishman}},
  \bibinfo{author}{\bibfnamefont{I.}~\bibnamefont{Guarneri}}, \bibnamefont{and}
  \bibinfo{author}{\bibfnamefont{L.}~\bibnamefont{Rebuzzini}},
  \bibinfo{journal}{Phys. Rev. A} \textbf{\bibinfo{volume}{73}},
  \bibinfo{pages}{052110} (\bibinfo{year}{2006}).

\bibitem[{\citenamefont{Mouchet}(2007)}]{Mou2007}
\bibinfo{author}{\bibfnamefont{A.}~\bibnamefont{Mouchet}}, \bibinfo{journal}{J.
  Phys. A} \textbf{\bibinfo{volume}{40}}, \bibinfo{pages}{F663}
  (\bibinfo{year}{2007}).

\bibitem[{\citenamefont{Shudo et~al.}(2002)\citenamefont{Shudo, Ishii, and
  Ikeda}}]{ShuIshIke2002}
\bibinfo{author}{\bibfnamefont{A.}~\bibnamefont{Shudo}},
  \bibinfo{author}{\bibfnamefont{Y.}~\bibnamefont{Ishii}}, \bibnamefont{and}
  \bibinfo{author}{\bibfnamefont{K.~S.} \bibnamefont{Ikeda}},
  \bibinfo{journal}{J. Phys. A} \textbf{\bibinfo{volume}{35}},
  \bibinfo{pages}{L225} (\bibinfo{year}{2002});
%
  \bibinfo{journal}{Europhys. Lett.} \textbf{\bibinfo{volume}{81}},
  \bibinfo{pages}{50003} (\bibinfo{year}{2008});
%
  \bibinfo{journal}{J. Phys. A} \textbf{\bibinfo{volume}{42}},
  \bibinfo{pages}{265101} (\bibinfo{year}{2009}{\natexlab{a}});
%
  \bibinfo{journal}{J. Phys. A} \textbf{\bibinfo{volume}{42}},
  \bibinfo{pages}{265102} (\bibinfo{year}{2009}{\natexlab{b}}).

\bibitem[{\citenamefont{B\"acker
  et~al.}(2008{\natexlab{b}})\citenamefont{B\"acker, Ketzmerick, L\"ock, and
  Schilling}}]{BaeKetLoeSch2008}
\bibinfo{author}{\bibfnamefont{A.}~\bibnamefont{B\"acker}},
  \bibinfo{author}{\bibfnamefont{R.}~\bibnamefont{Ketzmerick}},
  \bibinfo{author}{\bibfnamefont{S.}~\bibnamefont{L\"ock}}, \bibnamefont{and}
  \bibinfo{author}{\bibfnamefont{L.}~\bibnamefont{Schilling}},
  \bibinfo{journal}{Phys. Rev. Lett.} \textbf{\bibinfo{volume}{100}},
  \bibinfo{pages}{104101} (\bibinfo{year}{2008}{\natexlab{b}});
%
\bibinfo{author}{\bibfnamefont{A.}~\bibnamefont{B\"acker}},
  \bibinfo{author}{\bibfnamefont{R.}~\bibnamefont{Ketzmerick}},
  \bibnamefont{and} \bibinfo{author}{\bibfnamefont{S.}~\bibnamefont{L\"ock}},
  \bibinfo{journal}{Phys. Rev. E} \textbf{\bibinfo{volume}{82}},
  \bibinfo{pages}{056208} (\bibinfo{year}{2010}).

\bibitem[{\citenamefont{L\"ock et~al.}(2010)\citenamefont{L\"ock, B\"acker,
  Ketzmerick, and Schlagheck}}]{LoeBaeKetSch2010}
\bibinfo{author}{\bibfnamefont{S.}~\bibnamefont{L\"ock}},
  \bibinfo{author}{\bibfnamefont{A.}~\bibnamefont{B\"acker}},
  \bibinfo{author}{\bibfnamefont{R.}~\bibnamefont{Ketzmerick}},
  \bibnamefont{and}
  \bibinfo{author}{\bibfnamefont{P.}~\bibnamefont{Schlagheck}},
  \bibinfo{journal}{Phys. Rev. Lett.} \textbf{\bibinfo{volume}{104}},
  \bibinfo{pages}{114101} (\bibinfo{year}{2010}).

\bibitem[{\citenamefont{Deunff and Mouchet}(2010)}]{DenMou2010}
\bibinfo{author}{\bibfnamefont{J.}~\bibnamefont{Le~Deunff}} \bibnamefont{and}
  \bibinfo{author}{\bibfnamefont{A.}~\bibnamefont{Mouchet}},
  \bibinfo{journal}{Phys. Rev. E} \textbf{\bibinfo{volume}{81}},
  \bibinfo{pages}{046205} (\bibinfo{year}{2010}).

\bibitem[{\citenamefont{Miller}(1974)}]{Mil1974}
\bibinfo{author}{\bibfnamefont{W.~H.} \bibnamefont{Miller}},
  \bibinfo{journal}{Adv. Chem. Phys.} \textbf{\bibinfo{volume}{25}},
  \bibinfo{pages}{66} (\bibinfo{year}{1974}).

\bibitem[{\citenamefont{Schulman}(2005)}]{Sch2005}
\bibinfo{author}{\bibfnamefont{L.~S.} \bibnamefont{Schulman}},
  \emph{\bibinfo{title}{Techniques and Applications of Path Integration}}
  (\bibinfo{publisher}{Dover, Minola, New York}, \bibinfo{year}{2005}).

\bibitem[{\citenamefont{Murray}(1974)}]{Mur1974}
\bibinfo{author}{\bibfnamefont{J.~D.} \bibnamefont{Murray}},
  \emph{\bibinfo{title}{Asymptotic analysis}} (\bibinfo{publisher}{Clarendon
  Press, Oxford}, \bibinfo{year}{1974}).

\bibitem[{\citenamefont{Chirikov}(1979)}]{Chi1979}
\bibinfo{author}{\bibfnamefont{B.~V.} \bibnamefont{Chirikov}},
  \bibinfo{journal}{Phys.~Rep.} \textbf{\bibinfo{volume}{52}},
  \bibinfo{pages}{263} (\bibinfo{year}{1979}).

\bibitem[{\citenamefont{Berry et~al.}(1979)\citenamefont{Berry, Balzas, Tabor,
  and Voros}}]{BerBalTabVor1979}
\bibinfo{author}{\bibfnamefont{M.~V.} \bibnamefont{Berry}},
  \bibinfo{author}{\bibfnamefont{N.~L.} \bibnamefont{Balzas}},
  \bibinfo{author}{\bibfnamefont{M.}~\bibnamefont{Tabor}}, \bibnamefont{and}
  \bibinfo{author}{\bibfnamefont{A.}~\bibnamefont{Voros}},
  \bibinfo{journal}{Ann. of Phys.} \textbf{\bibinfo{volume}{122}},
  \bibinfo{pages}{26} (\bibinfo{year}{1979}).

\bibitem[{\citenamefont{Bogomolny}(1992)}]{Bog1992}
\bibinfo{author}{\bibfnamefont{E.~B.} \bibnamefont{Bogomolny}},
  \bibinfo{journal}{Nonlinearity} \textbf{\bibinfo{volume}{5}},
  \bibinfo{pages}{805} (\bibinfo{year}{1992}).

\bibitem[{\citenamefont{Greene and Percival}(1981)}]{GrePer1981}
\bibinfo{author}{\bibfnamefont{J.~M.} \bibnamefont{Greene}} \bibnamefont{and}
  \bibinfo{author}{\bibfnamefont{I.~C.} \bibnamefont{Percival}},
  \bibinfo{journal}{Physica D} \textbf{\bibinfo{volume}{3}},
  \bibinfo{pages}{530} (\bibinfo{year}{1981}).

\bibitem[{\citenamefont{Percival}(1982)}]{Per1982}
\bibinfo{author}{\bibfnamefont{I.~C.} \bibnamefont{Percival}},
  \bibinfo{journal}{Physica D} \textbf{\bibinfo{volume}{6}},
  \bibinfo{pages}{67} (\bibinfo{year}{1982}).

\end{thebibliography}
\end{document}